# Fractional Modified Special Relativity

Hosein Nasrolahpour[*]

Department of Physics, Faculty of Basic Sciences,
University of Mazandaran,
P. O. Box 47416-95447, Babolsar, IRAN

Hadaf Institute of Higher Education, P.O. Box 48175 – 1666, Sari, Mazandaran, IRAN

**Abstract:**

Fractional calculus represents a natural tool for describing relativistic phenomena in pseudo-Euclidean space-time. In this study, Fractional modified special relativity is presented. We obtain fractional generalized relation for the time dilation.

## 1- Introduction

The Theory of Special Relativity was first published as a paper in 1905[1]. Recently, some modifications on this theory have been proposed. For example, in ref. [3] some aspects of nonlocal special relativity have been studied. Dissipative Lagrangian for a relativistic particle also has been discussed in ref. [4].

In more recent years attention has been attracted to the application of Fractional Calculus (FC) to special (and general) relativity [9-13]. Fractional calculus can be used to model the nonlocal dissipative phenomena in space (-time) [7, 8]. In particular FC represents a natural tool for describing relativistic phenomena in pseudo-Euclidean space-time [9, 11].

The aim of this short paper is to introduce fractional modified special relativity as a new nonlocal dissipative model for the special theory of relativity. In the following, fractional calculus [5, 6] is briefly reviewed in Sec. 2. Fractional generalized relation for the time dilation is presented in Sec. 3. Finally, in Sec. 4, we will present some conclusions and discussion.

## 2-Fractional calculus: Riemann-Liouvill integral operator

Fractional calculus is the calculus of derivatives and integrals with arbitrary order, which unify and generalize the notions of integer order differentiation and n-fold integration. In contrast whit the definition of the ordinary derivative and integration, the definition of the fractional ones is not unique. In fact, there exists several definitions including, Grünwald - Letnikov, Riemann - Liouville, Weyl, Riesz and Caputo for fractional derivative and integral. For the purpose of this study we use the left-sided Riemann-Liouville fractional integral operator. The left-sided Riemann -Liouvill fractional integral of order $\alpha$ of a function f (t) is defined by [5, 6]

$$^{RL}_{0}I^{\alpha}_{t}f(t) = \frac{1}{\Gamma(\alpha)}\int_{0}^{t}(t-\tau)^{\alpha-1}f(\tau)d\tau \qquad t>0, \; \alpha \in \mathbb{R}^{+} \qquad (1)$$

[*] E-mail address: hnasrolahpour@gmail.com ; h.nasrolah-pour@umz.ac.ir

Where, $\mathbb{R}^+$ is the set of positive real numbers and $\Gamma(\alpha)$ *denotes the gamma function.

### 3-Time dilation for fractional special relativity

We start with the following expression for the fractional generalized position

$$x(t) = {}^{RL}_{0}I^{\alpha}_{t}(v_{\alpha}) \qquad\qquad 0 < \alpha \leq 1 \qquad\qquad (2)$$

By use of the above equation for the fractional generalized position we can easily derive dilated time formula according to the famous light clock experiment [2] as below

$$t = \frac{\sqrt{({}^{RL}_{0}I^{\alpha}_{t}(v_{\alpha}))^2 + (ct_0)^2}}{c} \qquad (3)$$

Where c is the speed of light in vacuum and $t_0$ is proper time. By use of Eq. (1), we can obtain the following relationship between $t_0$ (proper time) and t (dilated time):

$$t^2 - \frac{\beta^2 t^{2\alpha}}{\alpha^2 \Gamma^2(\alpha)} = t_0^2 \qquad (4)$$

We can write Eq. (4) as

$$t^2 - \frac{\beta^2 t^{2\alpha}}{\Gamma^2(\alpha+1)} = t_0^2 \qquad (5)$$

where $\beta = \frac{v_{\alpha}}{c}$. Now we can see that as $\alpha \to 1$, Eq. (4) gives

$$t^2 - \beta^2 t^2 = t_0^2 \qquad (6)$$

this leads to well-known time dilation formula of ordinary special relativity:

$$t = \frac{t_0}{\sqrt{1-\beta^2}}. \qquad (7)$$

If we plot $t$ (dilated time) with respect to $t_0$ for the special case of $v = 0.99c$ we will have

---

* Definition of Gamma function is given by $\Gamma(\alpha) = \int_0^{\infty} e^{-t} t^{\alpha-1} dt, \alpha \in \mathbb{R}^+$. For this function we have $\Gamma(\alpha+1) = \alpha\Gamma(\alpha)$.

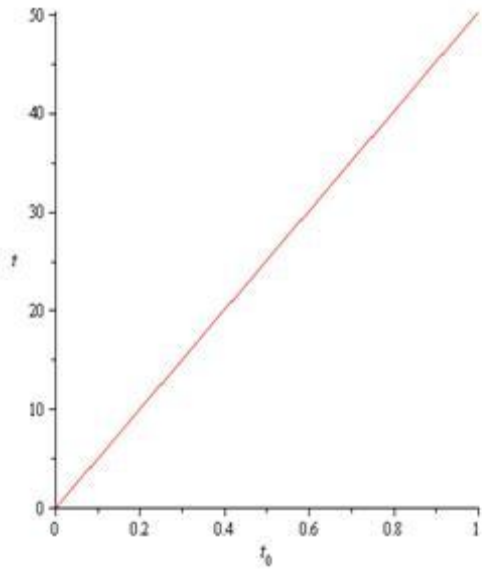

Fig. 1. $t$ with respect to $t_0$ for the special case of $v = 0.99c, \alpha = 1$

With the aid of expression (5) we can find the t (dilated time) versus $t_0$ (proper time) for the arbitrary case of $\alpha$. In particular, for the special case of $\alpha = \dfrac{1}{2}$ we can draw $t$ versus $t_0$

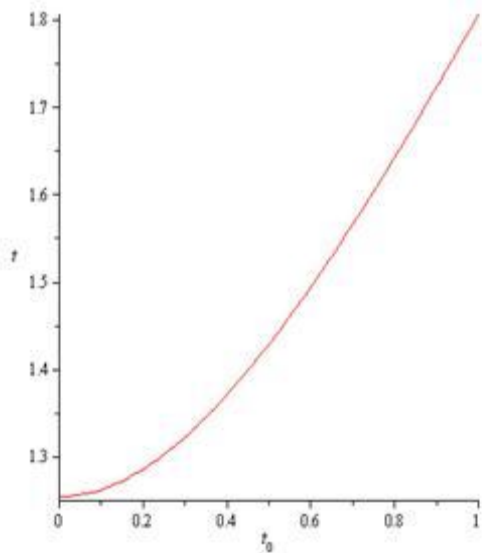

Fig. 2. $t$ with respect to $t_0$ for the special case of $v = 0.99c, \alpha = \dfrac{1}{2}$

As shown in this figure, there is a major difference between Fig.1 and Fig.2 near the origin of coordinate system. We can interpret this anomalous behaviour as an immediate consequence of applications of fractional calculus in our model.

## 4- CONCLUSION AND DISCUSSION

Fractional calculus can be used to model the nonlocal dissipative phenomena in space (-time). In particular FC represents an effective tool for describing relativistic phenomena in pseudo-Euclidean space-time. With this motivation we obtain fractional generalized relation for the well-known time dilation formula, to show the possibilities for future studies on fractional modified special theory of relativity theory. As we shown in Fig. (2), there is a anomalous behaviour when we plot $t$ with respect to $t_0$ for the arbitrary case of $\alpha$ ( $0 < \alpha < 1$ ) near the origin of coordinate system. We hope to study some other aspects of Fractional modified special relativity in future.